\documentclass[twocolumn,twocolappendix]{openjournal}
\usepackage{float}
\usepackage{xcolor,bm}
\usepackage{amsmath}

\usepackage{siunitx}

\begin{document}


\title{Probing Stellar Kinematics with the Time-Asymmetric Hanbury Brown and Twiss Effect}

\author{Lucijana Stanic$^1$, Ivan Cardea$^2$, Edoardo Charbon$^2$, Domenico Della Volpe$^3$, Daniel Florin$^1$, Andrea Guerrieri$^4$, Gilles Koziol$^3$, Etienne Lyard$^3$, Nicolas Produit$^3$, Aramis Raiola$^3$, Prasenjit Saha$^1$, Vitalii Sliusar$^3$, Achim Vollhardt$^1$, Roland Walter$^3$} 

\affiliation{$^1$University of Zurich, Switzerland}
\affiliation{$^2$École Polytechnique Fédérale de Lausanne (EPFL), Switzerland}
\affiliation{$^3$University of Geneva, Switzerland}
\affiliation{$^4$University of Applied Sciences and Arts Western Switzerland (HES-SO), Switzerland}

\begin{abstract}
    Intensity interferometry (II) offers a powerful means to observe stellar objects with a high resolution. In this work, we demonstrate that II can also probe internal stellar kinematics by revealing a time-asymmetric Hanbury Brown and Twiss (HBT) effect, causing a measurable shift in the temporal correlation peak away from zero delay. We develop numerical models to simulate this effect for two distinct astrophysical scenarios: an emission-line circumstellar disk and an absorption-line binary system. Our simulations reveal a clear sensitivity of this temporal asymmetry to the system's inclination angle, velocity symmetry, and internal dynamics. This suggests that, with sufficiently high time resolution, II can be used to extract quantitative information about internal kinematics, offering a new observational window on stellar dynamics.
\end{abstract}

\maketitle
 
\section{Introduction} \label{sec:introduction}
Intensity Interferometry (II) was first pioneered by Hanbury Brown and Twiss (HBT) in the 1950s \citep{HBT1956}, providing measurements of stellar diameters with an accuracy that reaches fractions of a milliarcsecond in the visible \citep{OG1974}.
This represented the highest angular resolution achieved in the optical spectrum at the time.
After a hiatus of almost half a century, primarily due to limitations in available technology, II has experienced a resurgence in astronomy.
One major catalyst for this revival, alongside advances in photon detection technology, was the construction of telescopes with large light-collecting areas and long baselines provided by Imaging Air Cherenkov Telescopes designed for gamma-ray astronomy.
Collaborations such as MAGIC \citep{magic2024}, VERITAS \citep{veritas2024} and H.E.S.S. \citep{hess2024} have successfully measured stellar diameters using II for the first time since the Narrabri Intensity interferometer was used by HBT.
But even simpler setups, such as small aperture telescopes equipped with detectors offering high temporal resolution, namely single-photon avalanche photodiodes (SPADs), have enabled the determination of position angles and axial ratios for extended disks surrounding stars \citep{Nolan2023}.
In parallel, large-area observational concepts have emerged. 
Notably, the QUASAR project \citep{walter2023} advocates for intensity interferometry systems using telescopes with mirror areas up to $\sim$ 1000 m$^2$ combined with ultra-fast photon counting via SPADs. 
This approach promises optical angular resolutions at the microarcsecond scale, allowing direct imaging of accretion disks and jets around compact galactic and extragalactic sources.

Unlike Michelson interferometry, which relies on wavefront coherence to reconstruct images, II exploits statistical correlations between intensity fluctuations, therefore not requiring phase-stable beam transport. 
This makes II a second-order technique, relying on correlations of intensities (second-order coherence) rather than electric field correlations (first-order coherence).
The fundamental idea is that the degree of correlation of photon counts between two distant detectors depends on the spatial distribution of the source's brightness.
Mathematically, this correlation can be expressed as a function that relates intensity fluctuations at different spatial positions, leading to a formulation of complex visibility.

In particular, consider a distant extended object, like a star, emitting light at wavelength $\lambda$.  
Let $(x,y, 0)$ denote the source-plane coordinates while $(u,v,d)$ denotes observer-plane coordinates with $d$ being the distance between the two planes.  
The complex visibility function that describes the correlations between the two points $(u_i,v_i)$ and $(u_j,v_j)$ in the observer plane, is given by the following integral expression:
\begin{align}
    &\Gamma(u_i,v_i;u_j,v_j) =
    \nonumber\\
    &\iint_{\text{source}} e^{-\frac{2\pi i}{\lambda} 
    \left(\frac{u_i-u_j}{d}x + \frac{v_i-v_j}{d}y\right)}
    \, I(x,y) \; dxdy
\label{eq:oldGamma}
\end{align}
with $|\Gamma|^2$ being the main observable in the case of intensity interferometry. 

Although equation \ref{eq:oldGamma} provides a way to quantify spatial correlations, it overlooks a wealth of information encoded in the dynamics of astrophysical systems.
In many objects, such as circumstellar disks around Be stars or orbiting binary systems, the emission regions exhibit ordered dynamical velocity fields.
In this paper we investigate how these internal kinematics manifest as a time-asymmetric HBT effect.
The approach of using intensity correlations to study dynamic sources, while novel in this specific astrophysical application, has a strong precedent in other fields.
In high-energy physics, the technique is known as femtoscopy and serves as a primary tool for measuring the spatio-temporal evolution of short-lived, expanding particle sources created in relativistic heavy ion collisions (for a technical review, see \cite{AnnanLisa2005}; for a historical overview, see \cite{Padula2005}).
The present work adapts this core concept to the astronomical domain with a new focus on extracting the signature of ordered internal kinematics from the temporal correlation function.

The physical signature of such an effect is governed by two distinct physical scales.  First, the coherence time $\Delta\tau$ is determined by the spectral bandwidth $\Delta \lambda$ of the light observed at the wavelength $\lambda$, according to
\begin{equation}
    \Delta\tau = \frac{\lambda^2}{c \, \Delta \lambda}
    \label{eq:cohtime}
\end{equation}
This $\Delta\tau$ defines the overall temporal width of the HBT peak.
Second, a kinematic signature arises from the Doppler shift induced by a LoS velocity $v$ with
\begin{equation}
    \frac{\Delta \lambda_{\rm kin}}{\lambda} = \frac{v}{c}
    \label{eq:doppler}
\end{equation}
Here we draw attention to a regime where the time-resolution is even better than the coherence time, thus
\begin{equation}
   \Delta t < \Delta\tau
\end{equation}
\textcolor{black}{while the Doppler shift in wavelength is comparable to the optical bandpass
\begin{equation}
   \Delta \lambda_{\rm kin} \sim \Delta\lambda
\end{equation}
On substituting for $\Delta\tau$ and $\Delta \lambda_{\rm kin}$ these two conditions can be rearranged as
\begin{equation} 
   \frac vc \; \sim \; \frac{\Delta\lambda}\lambda \; < \; \frac\lambda{c\,\Delta t}
   \label{eq:inequalities}
\end{equation}}%
With recently-developed photon detectors reaching $\Delta t \sim \SI{10}{\pico\second}$ at optical wavelengths, the outer inequality is satisfied for astrophysical sources at $v\sim\SI{10}{\kilo\metre\,\second^{-1}}$.  Satisfying the \textcolor{black}{condition on $\Delta\lambda$} then requires sub-nm bandwidths at optical wavelengths, which is feasible.  Exploration of this regime is therefore timely.

A key implication of the condition (\ref{eq:inequalities}) ---even when only approximately satisfied--- is that for source with rotation, the HBT correlation can become asymmetric in time, with an inverse relationship between the rotation velocity and the time asymmetry. 
Observing smaller LoS velocities at the same wavelength leads to a larger effect in the time domain, a counter-intuitive but powerful advantage for probing subtle kinematic regimes, in contrast to spectroscopy, where smaller velocities produce line shifts that are harder to resolve.

At this point it might be helpful to establish a more intuitive, physical picture of the phenomenon:
An extended, chaotic light source like a star does not project a smooth disk of light, but rather a complex and rapidly changing interference pattern into the far field of an observer.
The technique to study this so-called speckle pattern is referred to as speckle imaging and has been considered as an extension of interferometric methods \citep{Labeyrie1970}.
Intensity interferometry can be understood as correlating the intensity of this pattern at the two locations with two or more telescopes.
For a static source, this speckle pattern scintillates in place, leading to a symmetric temporal correlation.
When a source has an ordered velocity gradient, such as rotation, the Doppler shifts from the approaching and receding sides cause the entire speckle pattern to translate or drift across the observer's plan.
The kinematics of the source plane therefore manifest as a motion in the observer's speckle plane.
This drift explains the time-asymmetric HBT effect.
To find the maximum correlation between two detectors, one must introduce a processing time lag $\Delta t$, equal to the speckle's travel time between them.
Reversing the source's rotation reverses the drift direction thus requiring the opposite time lag, $\Delta t$, for the same baseline.

This physical picture is formalised by extending the visibility function to incorporate the Line-of-Sight (LoS) velocity distributions:
\begin{align}
  &\Gamma(u_i,v_i,t_i;u_j,v_j,t_j) = \nonumber \\
  &\iint_{\text{source}} e^{-\frac{2\pi i}\lambda \left(\frac{u_i-u_j}{d}x + \frac{v_i-v_j}{d}y + v_z(t_i - t_j) \right)} \, I(x,y) \; dxdy \nonumber \\
  &\quad
    \label{eq:newGamma}
\end{align}
In Appendix \ref{sec:app_deriv}, we provide a heuristic derivation of this formula from first principles, starting with the electric field of the source.
This work develops numerical models to simulate the LoS velocity distributions of circumstellar disks and binary systems, applying a Fourier analysis in three dimensions to study the effect of this incorporation of the dynamical effects of the system into the visibility function.

\section{Methods} \label{sec:methods}
To investigate the effect of internal kinematics on II, we developed a general framework for simulating a celestial light source and calculating its second-order correlation function.
This approach moves beyond the traditional static source assumption by incorporating the temporal dimension linked to the source's LoS velocities.
\subsection{Spatio-Temporal Visibility Function} \label{ssec:spacetimeV2}
The foundation of our work is an extension of the classical visibility function to include time-dependent phase information.
The quantum theory of optical coherence, formalised by \cite{glauber1963}, describes the full properties of a light field through correlation functions of spacetime points, $G^{(2)}(r_1,r_2,t_1,t_2)$.
The fact that the formalism is inherently four dimensional suggests that a source's temporal and spatial properties are intrinsically linked.

We propose a physically motivated model for the squared visibility that incorporates the first-order Doppler effect.
Instead of a continuous integral like in Equation \ref{eq:newGamma}, we model the source as a discrete collection of $N$ individual emitting particles.
The squared visibility function, as a function of the baseline vector components $(u,v)$ and a temporal delay $t$, is then computed as the squared modulus of the average complex phase exponential from all particles $k$
\begin{align}
    &|V_{12}(u,v,t)|^2 \nonumber \\
    &=\left| \left\langle \exp \left[ 2\pi i \left(\frac{u\cdot x_k + v\cdot y_k}{\lambda\,d} + \frac{v_{\text{LoS},k}}\lambda\,t \right) \right] \right\rangle_k \, \right|^2 
    \label{eq:visib}
\end{align}
This formula serves as the central tool for our investigation. 
The first term in the exponent, $(u\cdot x_i+v\cdot y_i)/(\lambda d)$, is the standard spatial phase term from classical interferometry, relating the baseline to the particle's projected position on the sky plane $(x_i, y_i)$. 
The second term, $(v_{\text{LoS},i} \cdot t)/\lambda$, is our crucial addition.
It introduces a kinematic phase that directly links the particle's LoS velocity, $v_{\text{LoS},i}$, to the temporal delay $t$. 
This term is what allows us to probe the source's internal dynamics.

\subsection{Numerical Particle Simulation}  \label{ssec:numsim}
To use Equation \ref{eq:visib}, we first construct a numerical model of the source as a population of particles.
This is a general method that can be adapted to various astrophysical scenarios in subsequent chapters.
The fundamental steps are:

\begin{enumerate}
    \item \textbf{Particle Generation:} The source is modelled as a collection of $N$ particles. Each particle $k$ is assigned a position $(x,y,z)$ and a three dimensional velocity vector $(v_x,v_y,v_z)$ in the source's rest frame based on a chosen physical model (e.g. Keplerian dynamics).
    \item \textbf{System Orientation:} The entire system is then mathematically rotated to match the desired viewing angle. This is achieved by applying the rotation matrices corresponding to the system's inclination and if necessary the position angle to all particle position and velocity vectors. This step transforms the coordinates into the observer's frame while preserving the structure 
    \item \textbf{Projection to 2D Observables:} Finally, the 3D information from the oriented system is projected into 2D quantities that an observer would measure:
    \begin{itemize}
        \item The \textbf{sky positions} $(x_k,y_k)$ are the transverse components of the rotated position vectors scaled by the distance $d$. In this step, the source's depth along the LoS (the rotated $z$ coordinated) is collapsed, effectively treating the source as a projection onto a single sky plane
        \item The \textbf{LoS velocity} $v_{\text{LoS},k}$ is the component of the rotated velocity vector that lies along the observer's line of sight.
    \end{itemize}
\end{enumerate}
The output of this simulation is a catalog of $N$ particles, each with a defined sky position $(x_k,y_k)$ and a LoS velocity $v_{\text{LoS},k}$.
To account for small-scale turbulent variations and observational uncertainties, we introduce a random fluctuation in the LoS velocity on the order of 10$\%$.
This catalog provides all the necessary inputs to compute the squared visibility via Equation \ref{eq:visib}.
By looping over a range of delay times $\tau$, we can map out the full temporal intensity correlation function and study its shape and symmetry.

\section{Application I: Emission Line Kinematics in a Circumstellar Disk} \label{sec:emission_disk}
To demonstrate the framework implemented in Section \ref{sec:methods}, we first apply it to an emission-dominated astrophysical system: a Be star.
This characteristic emission originates from a circumstellar decretion disk, which is a disk formed from material ejected directly from the rapidly rotating B star.
Unlike protoplanetary disks, which form around young stars during their formation, these disks are formed by episodic mass loss events from the star itself.
The ejected gas is then excited by the star's internal energy, leading to the observed Balmer line emission.
Initially, these disks were primarily observed spectroscopically, but more recently both Michelson interferometry \citep{Touhami2013} and II techniques \citep{Nolan2023} have been used to measure the disk's extent and orientation.
These systems are ideal test cases as their characteristic emission lines are kinematically broadened by the high velocity orbital motion of the gas surrounding the star, providing a strong, velocity-resolved signal.
Our model is inspired by the well-studied Be star, $\gamma$ Cassiopeiae, hereinafter referred to as $\gamma$ Cas, one of the first stars ever observed with an emission line \citep{Bestat1866, gamCas2017} and becoming classified as a Be type star. 

\subsection{Simulating the Keplerian Disk} \label{ssec:kepdisc}

\begin{figure*}
    \centering
    \includegraphics[width=0.95\linewidth]{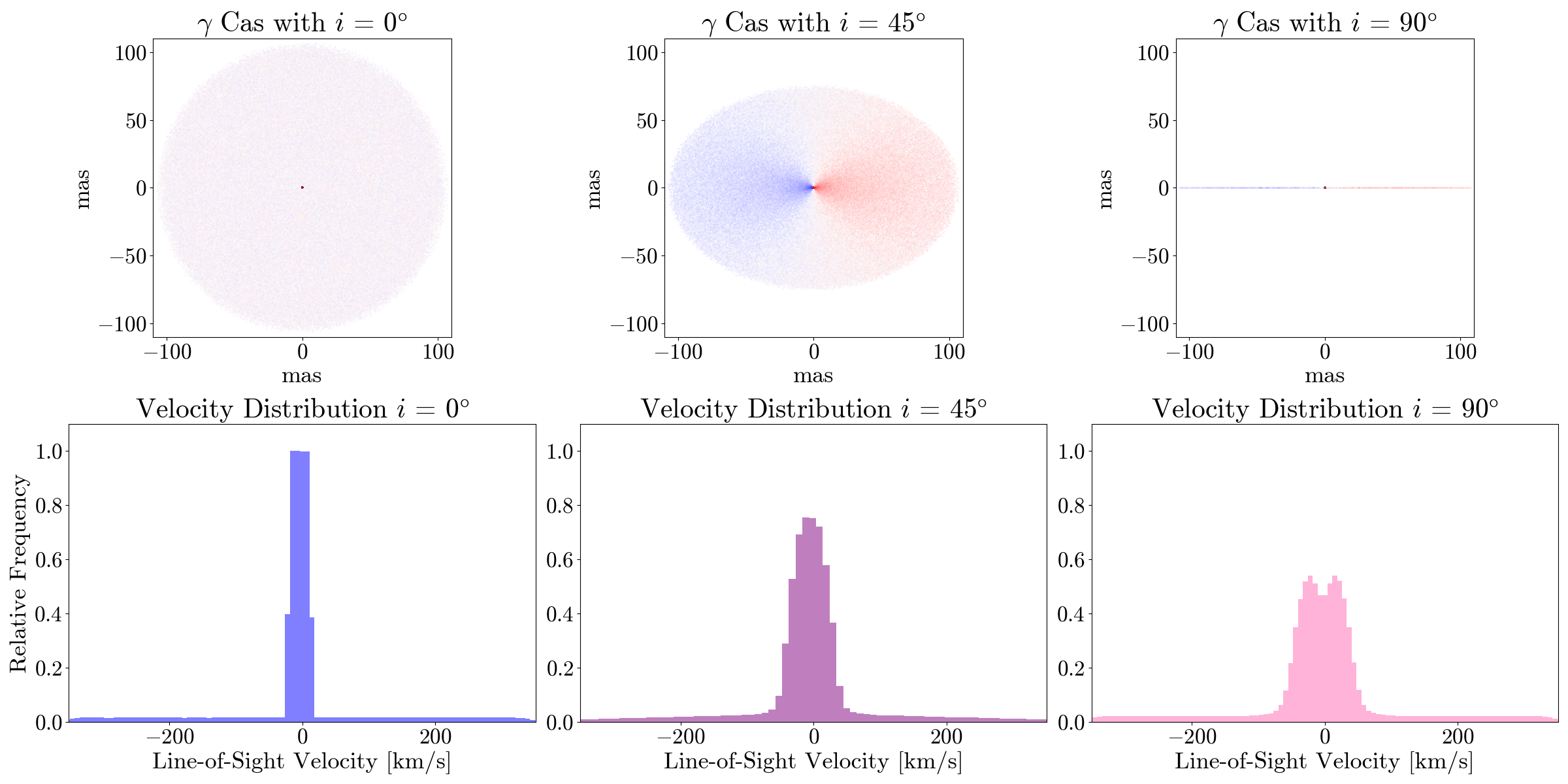}
    \caption{The Keplerian Disk modelled after the $\gamma$ Cassiopeia decretion disk is displayed here from three different viewing angles. A histogram of the line-of-sight (LoS) velocity distribution is presented for each of the views below. The distribution of velocities would correspond to a Doppler shift of the emission lines of up to $\pm$0.8nm}
    \label{fig:disks}
\end{figure*}

\begin{figure*}
    \centering
    \includegraphics[width=0.95\textwidth]{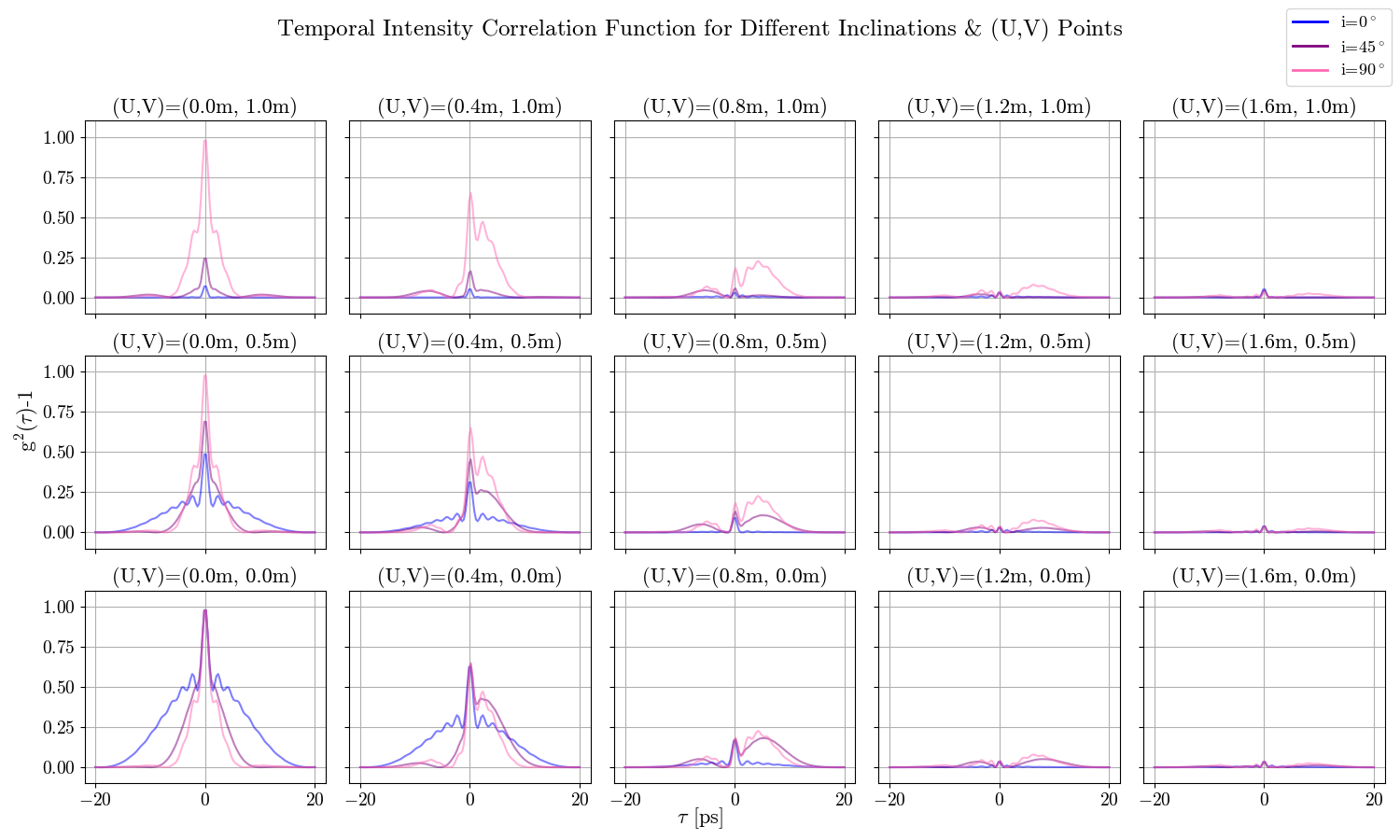} 
    \caption{The squared visibility as a function of delay time $\tau$ was plotted according to equation \ref{eq:visib}. In each plot one can see three distinct curves, color-coded according to the different inclination angles shown in Fig. \ref{fig:disks}. Each subplot corresponds to a certain point in the $(u,v)$-plane, of which only one corner is shown, since for negative $v$ we would see an exact copy of the positive parts and for negative $u$ we would observe a mirrored version along  the $u=0$ axis.}
    \label{fig:visibs}
\end{figure*}

The simulation follows the general steps outlined in Section \ref{ssec:numsim}, with the physics of a thin, axisymmetric Keplerian disk dictating the initial particle properties.
We populate the disk with $3\cdot 10^6$ particles, whose positions and velocities are determined by assigning each particle a Keplerian orbit around a central star with parameters taken from $\gamma$ Cas and listed in Table \ref{tab:disk}.
To create a realistic disk structure, semi-major axes are sampled uniformly in area between an inner and outer radius, and orbital eccentricities are assigned based on radial position.
The instantaneous positions are calculated using the standard orbital equations and velocities are initially approximated by the circular Keplerian velocity $v_{\text{circ}} = \sqrt{GM/r}$.
The 3D particle cloud representing the disk is then oriented to a specific viewing angle at selected inclination angles, from face-on ($i=0^{\circ}$), where the LoS is zero, to edge-on ($i=90^{\circ}$), where the full range of orbital velocities is projected along the LoS
The resulting disk viewed from three different angles and the accompanying LoS velocity distribution are depicted in Fig. \ref{fig:disks}.

\begin{table}
    \centering
    \caption{Parameters used to simulate a decretion disk similar to $\gamma$ Cas}
    \begin{tabular}{ll}  
        \hline
        Distance                & 168 pc \citep{survey2021}       \\ 
        Mass of central star    & 13 M$_\odot$ \citep{survey2021} \\ 
        Radius of central star  & 10 R$_\odot$ \citep{hipparcos1997} \\ 
        Outer Radius of disk    & 14 AU   \citep{stee2012}     \\ 
        Wavelength              & 656 nm       \\ 
        Particles               & 3 $\cdot$ 10$^6$       \\ 
        \hline
    \end{tabular}
    \label{tab:disk}
\end{table}

\subsection{The Role of the Spectral Filter in Emission} \label{ssec:emfilter}
The spectral filter is a fundamental component in any II measurement.
By isolating a quasi-monochromatic band of light, it defines the temporal coherence of the signal, which in turn determines the overall temporal width of the measured correlation function.
In an emission line scenario, the astrophysical source itself defines the spectral profile of the signal.
The primary role of an observational spectral filter is therefore to act as a bandpass filter, isolating the emission line of interest, in our case the 656 nm H$_\alpha$ line, and rejecting contaminating light from the stellar continuum or other spectral features.
As long as the filter's bandpass is wide enough to encompass the entire velocity broadened line, its specific transmission shape has a negligible impact on the resulting correlation function.
This is a fundamental distinction from the absorption line case discussed in the next chapter, where the filter's properties are paramount in defining the signal.

\subsection{Resulting Kinematic Signatures} \label{ssec:diskres}
\begin{figure*}
    \centering
    \includegraphics[width=0.95\linewidth]{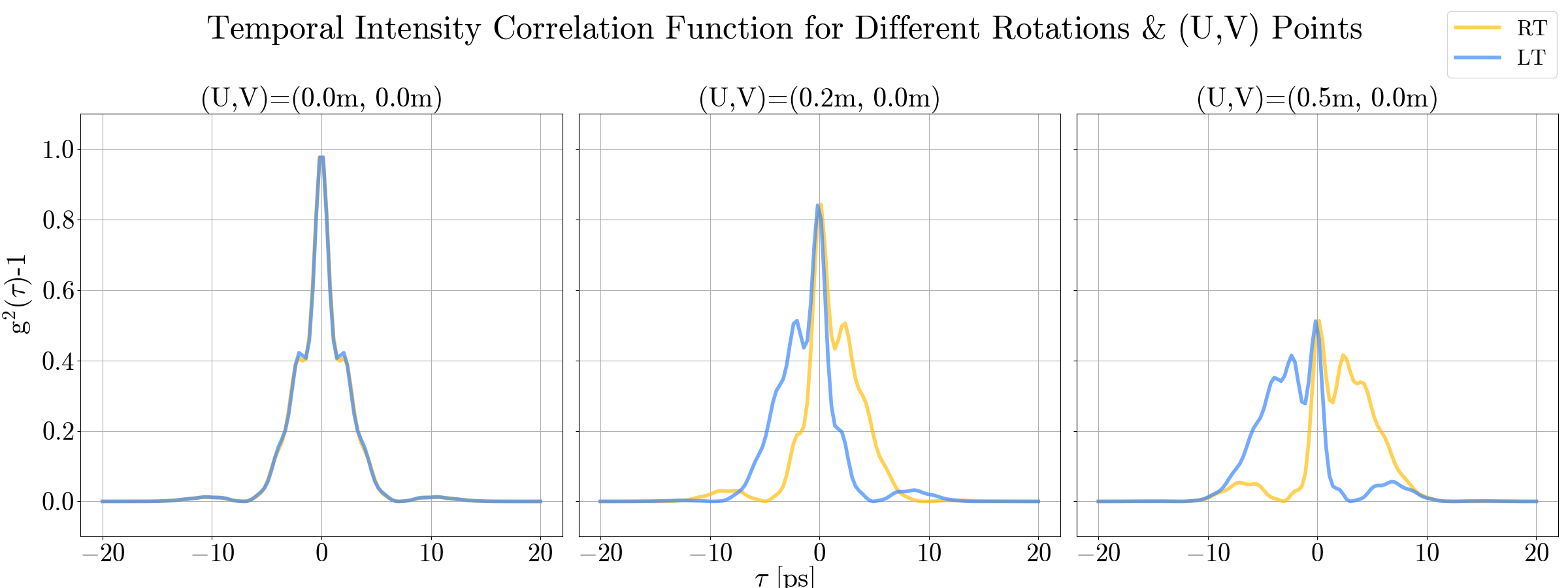}
    \caption{Comparison of the same disk viewed with an inclination of 90$^\circ$, but with two different rotations. We see that the peak shifts to a positive time delay for an increasing $u$ in the $u,v$-plane for a disk with right rotation, while it decreases at the same point if the disk is left rotating.}
    \label{fig:RTLT}
\end{figure*}

Applying our spatio-temporal visibility formula (Equation \ref{eq:visib}) to the simulated disk, reveals the kinematic dependencies:

\begin{enumerate}
    \item \textbf{Effect of Inclination:} As shown in Fig. \ref{fig:disks}, increasing the inclination from $0^\circ$ to $90^\circ$ broadens the distribution of LoS velocities. This has a direct effect on the temporal correlation function showcased in Fig. \ref{fig:visibs}. At a zero baseline the wider velocity distribution of the edge-on disk produces a correspondingly narrower correlation peak in the time domain, a direct consequence of the Fourier relationship between the spectrum and the temporal correlation.
    \item \textbf{Time-Asymmetric HBT Effect:} For any inclination greater than zero and a non-zero baseline, the symmetry of the correlation function is broken. Fig. \ref{fig:visibs} shows that the peak of the correlation function shifts away from $\tau =0$. This asymmetry is a direct measurement of the ordered velocity gradient across the disk as viewed along the specific direction of the baseline.
    \item \textbf{Sensitivity to Rotation Direction:} The sign of this time shift is directly tied to the disk's sense of rotation. As demonstrated in Fig. \ref{fig:RTLT}, for a given baseline, a right-turning disk produces a peak shift in one direction of $\tau$ while a left-turning disk produces a shift in the opposite direction. This confirms that the time-asymmetric HBT effect is sensitive not just to the magnitude of the velocities, but also to the global kinematic structure rotation.
\end{enumerate}

\section{Application II: Absorption Line Kinematics in a Binary System} \label{sec:binaries}
We now turn to an absorption dominated system to illustrate the versatility and the different nature of the time-asymmetric HBT in this regime.
A spectroscopic binary system serves as an excellent case study, where the kinematic signature is imprinted as Doppler-shifted absorption lines from the star's photospheres against the bright continuum.
Our model is inspired by the binary system $\gamma$ Persei, from this point onwards referred to as $\gamma$ Per.

\subsection{Simulating the Binary System} \label{ssec:simbin}

We model the binary system as two distinct spherical light sources in a Keplerian orbit.
The photospheres of the primary and secondary stars are each populated with particles.
All particles belonging to a single star are assigned the same orbital velocity vector, calculated from the system's known orbital parameters listed in Table \ref{tab:binary}.
The absorption line effect is simulated by creating a composite velocity distribution.
The continuum from both stars is modelled as a broad, flat distribution, while we assume a Gaussian filter response. 
Therefore, the absorption lines are simulated as two distinct Gaussian decrements superimposed as two distinct Gaussian decrements in this distribution, centered at the respective LoS velocities of the two stars.
The width of these Gaussian decrements corresponds to the intrinsic width of the stellar absorption lines.
As before, the system is oriented by varying inclination angle, which projects the orbital velocities onto the observer's LoS, determining the observed separation of the two absorption features.

\begin{table}[hbt]
    \centering
    \caption{Parameters used to simulate a binary system similar to $\gamma$ Per}
    \begin{tabular}{ll}  

        \hline
        Distance                    & 68 pc \citep{ling2001}       \\ 
        Mass of primary star        & 3.6 M$_\odot$ \citep{diamant2023} \\ 
        Radius of primary star      & 22.7 R$_\odot$ \citep{diamant2023} \\ 
        Mass of secondary star      & 2.4 M$_\odot$ \citep{diamant2023} \\ 
        Radius of secondary star    & 3.9 R$_\odot$ \citep{diamant2023} \\ 
        Semi-major axis              & 9.6 AU   \citep{ling2001}     \\ 
        Wavelength                  & 393 nm       \\ 
        Particles                   & 3 $\cdot$ 10$^6$       \\ 
        \hline
    \end{tabular}
    \label{tab:binary}
\end{table}

\subsection{The Central Role of the Spectral Filter in Absorption} \label{ssec:filterabs}
In contrast to the emission line case, where the source itself provides a spectrally-varying signal, an absorption line system presents a different challenge: The signal is a narrow absence of the light against a bright, spectrally flat continuum.
For a given broadband stellar flux, this inherently results in a lower signal-to-noise ratio than for a bright emission line that can rise well above the continuum.
To extract this subtle signal, the filter must act as a transducer, converting the Doppler shift of the absorption line into a measurable intensity change as the line moves across the filter's transmission bandpass.
The resulting signal is a direct convolution of the absorption line's profile with the filter's transmission profile.
Consequently, the filter's shape, whether a steep rectangular or a shallower Gaussian profile, is of importance.
In our model, we simulate this process by assuming an effective Gaussian filter response.

\subsection{Results: Kinematic Signatures in Absorption} \label{ssec:resbin}
\begin{figure*}
    \centering
    \includegraphics[width=0.95\linewidth]{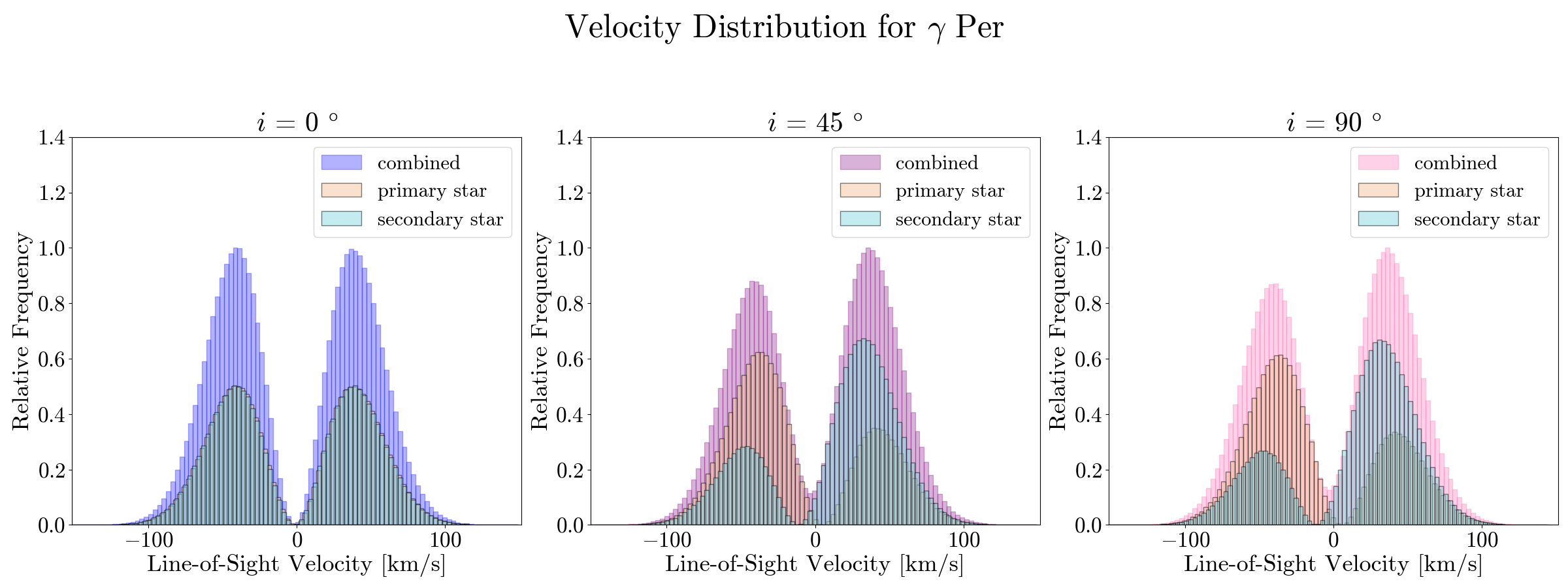}
    \caption{A histogram of the binary system's velocity distribution. The velocity distribution corresponds to observing the spectrum within a filter centered around a a wavelength with artificially added decreases representing the absorption lines. The differences between 45$^\circ$ and 90$^\circ$ for this model compared to the ones of the disk seen in Fig. \ref{fig:disks} seems negligible.}
    \label{fig:binvel}
\end{figure*}

\begin{figure*}
    \centering
    \includegraphics[width=0.95\linewidth]{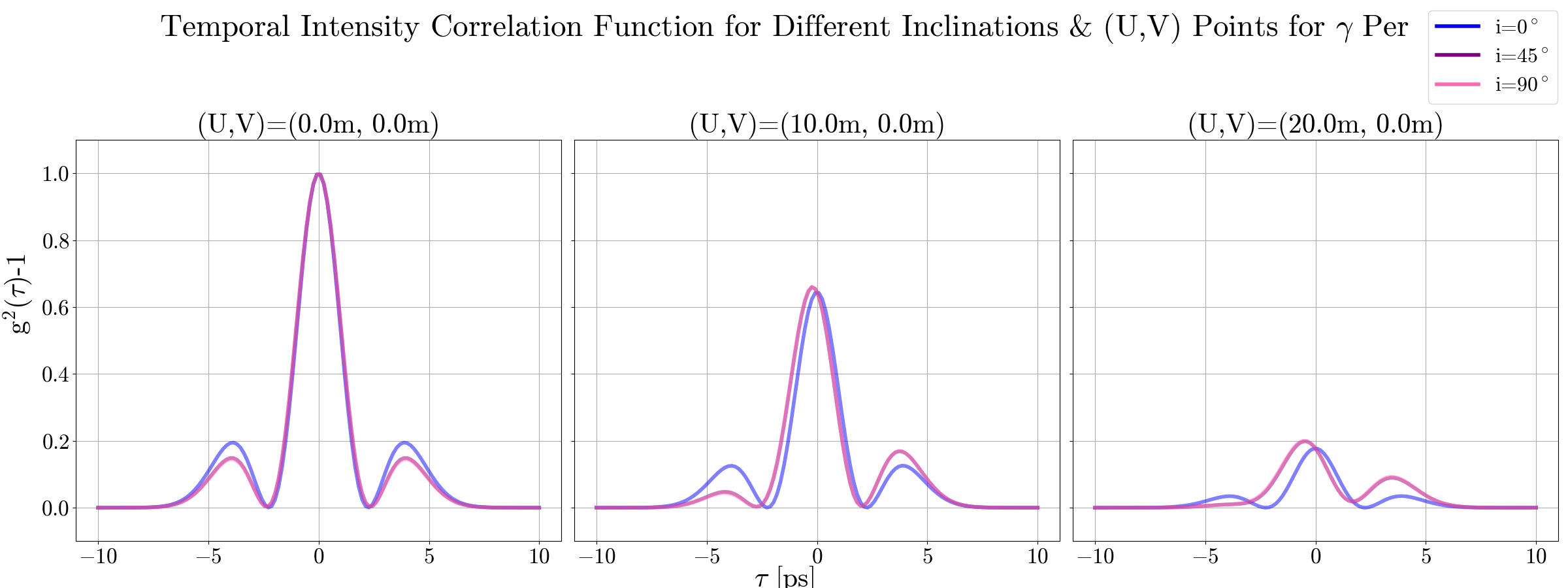}
    \caption{Similarly as in \ref{fig:visibs} the squared visibility of the binary functions as a function of delay time $\tau$ was plotted according to equation \ref{eq:visib}. Subplots correspond again to different points in the $u,v$-plane. The peak shape does not change as drastically with increasing $u$ and $v$ as it did for the previous model, but a clear asymmetry remains. The curves for 45$^\circ$ and 90$^\circ$ overlap almost completely. Points with non-zero $v$ are not shown, as they primarily exhibit a decrease in amplitude without significantly altering the temporal shape of the function.}
    \label{fig:bincorr}
\end{figure*}

The resulting velocity distributions depicted in Fig. \ref{fig:binvel} show two distinct dips corresponding to the absorption lines of the two stars.
Appyling our visibility formula to this model yields the following results:
\begin{enumerate}
    \item \textbf{Subtler Dependence on Inclination:} Unlike the extended disk, the binary system consists of two quasi-point sources. As a result, the correlation function's shape shows a subtler dependence on inclination, as seen in Fig. \ref{fig:bincorr}. The primary change with inclination is the velocity separation of the two absorption features, which is maximal at $i=90^\circ$.
    \item \textbf{Asymmetry from Orbital Motion:} Despite the simpler dynamics, a clear time-asymmetric HBT effect is still present for non-zero baselines and inclinations. The asymmetry in the correlation function arises from the velocity difference between the two stars, as probed by the directional baseline.
    \item \textbf{Less Complex Structure:} The correlation functions for the binary systems do not exhibit the same degree of structural change with increasing baseline as the disk model did. This is expected, as the velocity field is composed of two discrete components rather than the continuous, sheared velocity field of a rotating disk. Nonetheless, the presence of the asymmetry confirms that the technique is sensitive to the ordered motion in both emission and absorption regimes.
\end{enumerate}

\textcolor{black}{\section{Practical Limitations and Observational Feasibility}}\label{sec:practical_limits}
\textcolor{black}{The theoretical results presented in this work show that kinematic signatures can be imprinted on the temporal correlation function at the picosecond level. 
While these results are formally correct, it is crucial to discuss the practical limitations that would occur by an attempt to resolve this. 
The primary challenges that are up for discussion are atmospheric turbulence and photon statistics.}

\textcolor{black}{\subsection{The Effect of Atmospheric Turbulence}} \label{ssec:atmosphere}
\textcolor{black}{II is famously insensitive to the first-order phase fluctuations impairing traditional amplitude interferometry.
However, the relative arrival times of intensity patterns to observe the HBT effect can be affected by second-order path-length fluctuations.
\cite{HanburyBrown1975} conducted an analysis of the impact that the differential atmospheric path delay between two light paths would have.
They concluded that even in extreme cases, the fluctuations were unlikely to exceed 1 ps and would cause negligble loss of correlation.}

\textcolor{black}{The original prediction seems to stand in contrast of modern measurements from the field of Lunar Laser Ranging (LLR), which routinely measures timing jitter of about 10 ps \citep[e.g.,][]{Ortolani2011}.
This atmospheric noise floor would make it extremely challenging, if not impossible, to resolve the sub-10-picosecond asymmetries predicted in our simulations from the ground.
However, it is important to note that the LLR measurement represents a scenario involving a two-way path through the entire atmosphere, with a 2.6 second delay between outgoing and incoming light, during which the atmospheric column changes completely.
The differential jitter for an SII measurement, where two nearby telescopes view a star through highly correlated column of air is expected to be smaller, though it remains a fundamental noise source.
The $\approx$ 10 picosecond figure should therefore be considered a conservative upper limit for ground-base SII.}

\textcolor{black}{Consistent with this expectation, recent measurements of the HBT correlation peak of the Sun \citep{Sliusar2026} indicate an atmospheric broadening with an upper limit of approximately 6 ps (RMS), supporting the interpretation that the 10 ps LLR value represents a conservative bound for ground-based SII.}

\textcolor{black}{The limitation of atmospheric noise could be eliminated altogether by carrying out the observations from space-based instruments like proposed by \cite{Klein2007}, where atmospheric effects are absent.}

\textcolor{black}{\subsection{The Photon Flux Limit}} \label{ssec:SNRestimate}
\textcolor{black}{The second major challenge in observing the asymmetric effect pertains to the statistics of photons. 
We derive an estimate for the two objects for which the model was developed.}

\textcolor{black}{For $\gamma$ Cas the apparent magnitude in the R band centered close to the H$_\alpha$ line is 2.3 \cite{Wenger2000}.
This corresponds to a spectral photon flux density $\Phi$ of roughly $\Phi$ = 7 $\cdot$ 10$^{-6}$ photons m$^{-2}$ s$^{-2}$ Hz$^{-2}$ \textcolor{black}{in one polarization channel}.
The emission line has exhibited variability over the past few decades, where recent measurements \cite{BeSS2011} have shown the flux of the spectral line to be 6 times higher in intensity then the the surrounding continuum, therefore $\Phi$ = 4 $\cdot$ 10$^{-5}$ photons m$^{-2}$ s$^{-2}$ Hz$^{-2}$, same as a star with apparent magnitude 0.5.
The Signal-to-Noise Ratio (SNR) in SII can be estimated using
\begin{equation} \label{eq:SNR}
	\mathrm{SNR}=\Phi |V|^2 A \varepsilon \sqrt{\frac{T}{\Delta t}}
\end{equation}
where $A$ corresponds to the mirror area of one of the telescopes, $\varepsilon$ gives the optical and detector efficiency and $T$ the observational time. \textcolor{black}{(This formula often appears slightly differently: considering two polarization channels, and replacing $1/\Delta$ with twice the sampling frequency.)} Using telescopes with diameter of 1 meter with a total efficiency of 10 $\%$ and observing the spectral line of $\gamma$ Cas for an hour with $\Delta t$ of a picosecond would lead to a SNR of almost 200, with $\Delta t$ = 10 picosecond that would go down to SNR of roughly 60.}

\textcolor{black}{In the case of the absorption line, it is to be understood that the intensity would be reduced in comparison with the continuum. 
This would result in the necessity of longer observation times in order to achieve the same SNR. 
In particular, \citep{diamant2023} measured the spectral flux reduced to a quarter to the relative flux in the continuum going from magnitude 4.0 in the U band to 5.2, leading to SNR of about 2 with the same setup described above with the more optimistic case of $\Delta t$ = 1 ps.
This makes the absorption-line measurements significantly more challenging and restricts the feasibility to brighter targets or larger telescopes.}

\textcolor{black}{\subsection{Parameter Choices and Physical Regimes}} \label{ssec:fundamental}
\textcolor{black}{Crucially, the $\sim$ 10 ps atmospheric timing jitter is not a fundamental showstopper for the technique.
As established in Section \ref{sec:introduction}, the magnitude of the temporal asymmetrx is inversely proportional to velocity.
Observation of a slower-moving source will result in a larger and more easily measurable time delay, provided that the spectral filter width is sufficiently small.
A source moving at 10 km/s would for example lead to a time delay of orders of tens of picoseconds using a spectral filter with a bandwidth of 0.02 nm.}

\textcolor{black}{This shows that the atmospheric limit does not invalidate the technique but rather focuses its application on a specific regime. 
The higher-velocity regimes and sub-picosecond asymmetries explored in our simulations represent the ideal, atmosphere-free case that could be accessible to a future space-based intensity interferometer, while the larger shifts from lower-velocity sources remain a viable target for ground-based observatories.}
\newpage

\section{Discussion and Conclusion} \label{sec:disc_con}
Our simulations demonstrate that the temporal intensity correlation function is sensitive to the internal kinematics of a light source. 
We have shown that an ordered, asymmetric LoS velocity distribution, such as that produced by rotation, manifests itself as a time-asymmetric HBT effect.
This results in a temporal correlation peak that is skewed or shifted away from a time delay of $\tau=0$ s.
This work establishes a theoretical basis using this asymmetry as a new observable for probing the dynamics of astrophysical systems.

We have explored this effect in two distinct scenarios: an emission-dominated circumstellar disk and an absorption-dominated binary star system.
Despite the different physical mechanisms for signal generation, the interaction between the directional baseline of the interferometry setup and the source velocity gradient produces in both cases an asymmetry in the time domain.
The shape and direction of this asymmetry are directly linked to the geometry and internal motion of the system, such as its inclination angle and direction of rotation.

\begin{figure*}
    \centering
    \includegraphics[width=0.95\linewidth]{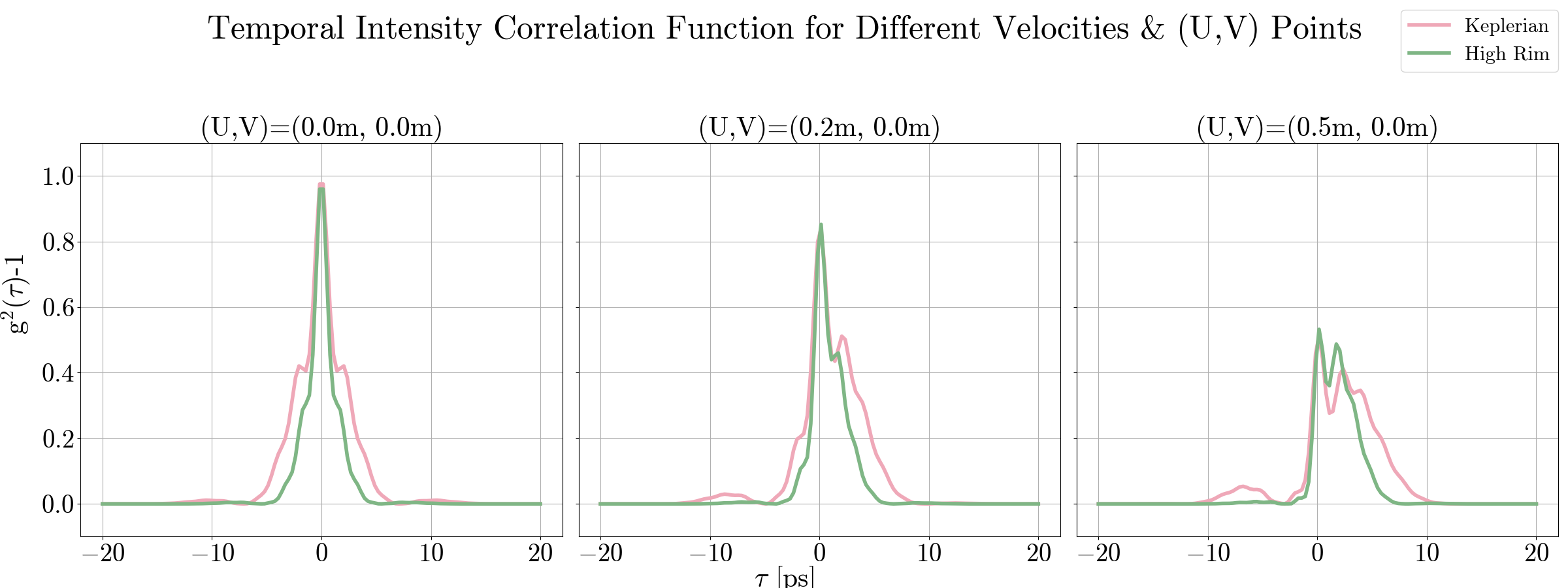}
    \caption{Comparison between two disks with different velocity model, viewed from an inclination angle of 90 $^\circ$. The pink one shows the classical Keplerian disk, the same as the ones shown in previous plot above. The green one shows a disk with the velocities increasing with distance to the center.}
    \label{fig:keprim}
\end{figure*}

The primary contribution of this work is the prediction that, if observed, this time-asymmetric HBT effect could become a new method for measuring kinematics in the time domain.
A key advantage of this technique is its powerful sensitivity to low-velocity fields.
As demonstrated by the relation $v\tau \sim \lambda$, measuring at the same wavelength, the impact of a smaller velocity $v$ will be compensated by a larger, easier to measure time delay $\tau$.
This is in contrast to high-resolution spectroscopy, where a smaller velocity always produces a smaller, harder-to-measure signal.
The generality of this effect is further illustrated by our simple example of a disk with non-Keplerian velocities showcased in Fig. \ref{fig:keprim}.
While not intended to represent a specific real-world object, this model shows that the kinematic signature is a fundamental consequence of the velocity field's structure.

It is important to note, however, that this paper focused exclusively only on the predictive modeling task, namely simulating an observable effect from a known physical model.
The complementary inference task, which involves deriving the underlying velocity field from an observed asymmetric correlation function, has not been studied here and remains a critical next step for practical application of this technique.

These theoretical predictions also suggest a path toward a laboratory validation. 
Such experiments would build upon strong foundation of recent work in creating and correlating pseudo-thermal light in tabletop settings to demonstrate the HBT effect, e.g. \cite{Dravins2015} and \cite{NituRai2024}.
Previous photon-correlation spectroscopy studies have already illustrated the feasibility of detecting narrow-line astrophysical signals in controlled laboratory conditions \citep{DravinsGermana2008}.
Its astrophysical applications, from detecting natural lasers to probing coherence properties of light, are reviewed in detail by \cite{Dravins2008}.
These preceding experiments have largely focused on symmetric temporal correlations arising from broadening caused by thermal motion in scattering fluids.
A direct test of our prediction, however, requires demonstrating the asymmetry that arises from an ordered velocity gradient.
A more effective experiment design would be to create two pseudo-thermal sources with slight, stable wavelength offset, which serves as a direct and controllable analogue to a kinematic Doppler shift.
This could be achieved, for example by passing monochromatic light though a dispersive element, be it prism or diffraction grating to induce a small calibrated wavelength shift.
An even more direct method would involve using two tunable lasers, precisely locked to slightly different frequencies.

In conclusion, this work establishes a theoretical foundation for using temporal II as a probe of stellar kinematics.
With the advent of single-photon detectors pushing towards picosecond time resolution (see for example \cite{Gramuglia2022, Hao2024}), observing the time-asymmetric HBT effect is becoming technologically feasible.
Indeed, this is precisely the domain that future astronomical instruments are being designed to explore.
Conceptual design studies for quantum-optics instruments on Extremely Large Telescopes, such as the QuantEYE concept for the former ESO OWL project, were predicated on exactly such detector capabilities to perform photon correlation spectroscopy and explore novel coherence effects \citep{Naletto2006}.
This could potentially open a new window onto the dynamics of a wide range of astrophysical systems, providing quantitative insights that are complementary to existing observational methods.

\section*{Acknowledgments}
We would like to thank Dainis Dravins for providing several key literature pointers that greatly assisted in situating this work in its broader scientific context.

\textcolor{black}{Additionally we would like to thank the anonymous Referee for their helpful comments and feedback, which greatly strengthened the overall manuscript.}

\textcolor{black}{This research has made use of the SIMBAD database, CDS, Strasbourg Astronomical Observatory, France and of the BeSS database, operated at LESIA, Observatoire de Meudon, France: http://basebe.obspm.fr}

This work was supported by the Swiss National Science Foundation (SNSF) through the Sinergia grant “Resolving Accretion Disks with Quantum Optics” (grant no. 216669).
\newpage
\bibliographystyle{aa}
\bibliography{references}

\onecolumngrid

\appendix
\section{Derivation of the Spatio-Temporal Coherence Function} \label{sec:app_deriv}

Consider the electric field $E(u,v,t)$ measured in the observer's plane $(u,v)$ at a time $t$.  This can be described by the superposition of the fields emitted from all points $(x,y)$ from the source as followed:
\begin{equation}
    E(u,v,t) = \iint_{\text{source}} A\left(x,y, \nu \right) \frac{e^{-2\pi i \nu \left( t - \frac{R}{c}\right)}}{R} dxdy
\end{equation}
where $A(x,y,\nu)$ represents the complex amplitude of the light emitted from the source at a position $(x,y)$ and $\nu$ is the frequency of light.
The distance $R$ from an infinitesimal source element to the detection point $(u,v,d)$ is
\begin{equation}
    R = \sqrt{\left( (u-x)^2 + (v-y)^2 + d^2\right)} = d\sqrt{\left(1 + \frac{(u-x)^2 + (v-y)^2}{d^2} \right)} \approx d\left( 1 + \frac{(u-x)^2 + (v-y)^2}{2d^2} \right)
\end{equation}
Here, we have used a binomial approximation. Expanding the square leads to
\begin{equation}
    R =  d\left( 1 + \frac{(u-x)^2 + (v-y)^2}{2d^2} \right) \approx d + \frac{u^2+v^2}{2d} +\frac{x^2+y^2}{2d} - \frac{ux+vy}{d}
\end{equation}
Subtracting the $d$ to have the phase at the measuring point be relative to the phase at $(0,0)$ and substituting this into the integral for the electric field
\begin{equation}
    E(u,v,t) =  \iint_{\text{source}}A(x,y,\nu) e^{-2\pi i \nu  \left(t - \frac{u^2+v^2}{2cd} - \frac{x^2+y^2}{2cd} + \frac{ux+vy}{cd}\right)} dxdy
\end{equation}
The $1/R$ amplitude factor varies slowly compared to the rapid oscillations for the exponential term, and can therefore be approximated as a constant and neglected in the integral. 
Furthermore we factor in the exponentials containing the dependencies on only $x$ and $y$ into the complex amplitude $A(x,y,\nu)$, since it is not directly measurable and can include such phase factors.
Lastly, we also apply the far-field (Fraunhofer) approximation to the phase terms.
This is valid because the distance to the source $d$ is vastly larger than both the source's physical size $(x,y)$ and the observing baselines $(u,v)$.
This approximation allows us to neglect the quadratic terms, those proportional to $u^2$ and $v^2$ in the full expansion of the path length as they contribute neglibly to the linear cross terms.
The phase of the electric field integral is thus simplified to
\begin{align}
    E(u,v,t) &= \iint_{\text{source}} A(x,y,\nu) e^{-2\pi i \nu \left(t + \frac{ux+vy}{cd}\right)} dxdy
\end{align}
Now the mutual coherence function or correlation function, which describes how well the field $E(x,y,t)$ at one space-time point $(u_i,v_i,t_i)$ correlates with the field at another one at $(u_j,v_j,t_j)$, is
\begin{align}
    \Gamma(u_i,v_i,t_i;u_j,v_j,t_j) &= \langle E(u_i,v_i,t_i)E^*(u_j,v_j,t_j)\rangle  \nonumber \\
    = \left\langle 
    \iint_{\text{source}} A(x,y,\nu) e^{-2\pi i \nu \left(t_i + \frac{u_ix+v_iy}{cd}\right)} dxdy \right.  &\times
    \left.\iint_{\text{source}}A(x',y',\nu) e^{+2\pi i \nu \left(t_j + \frac{u_jx'+v_jy'}{cd}\right)} dx'dy'
    \right\rangle
\end{align}
With the assumption that the source field $A(x',y',\nu)$ is spatially incoherent, we introduce the correlation function of $A(x,y,\nu)$
\begin{equation}
    \langle A(x,y,\nu)A^*(x',y',\nu)\rangle = I(x,y,\nu) \delta(x-x')\delta(y-y')
\end{equation}
which simplifies the above integral to
\begin{equation}
    \Gamma(x_i,y_i,t_i;x_j,y_j,t_j) = \iint_{\text{source}}I(x,y,\nu) e^{-2\pi i \nu \left(t_i - t_j + \frac{u_i-u_j}{cd}x + \frac{v_i-v_j}{cd}y\right)} dxdy
    \label{eq:Gamma_app1}
\end{equation}

This equation describes the mutual coherence function for a static, quasi-monochromatic source. 
The relationship between the source's spatial intensity distribution, $I(x,y)$, and the spatial part of the coherence function is formally known as the van Cittert-Zernike theorem. 
The standard theorem, however, is formulated for a static source and does not account for internal kinematics. 
The temporal phase term, $\exp{(-2\pi i\nu(t_i - t_j))}$, is treated as a simple oscillation at a single frequency.
To probe kinematics, we must extend this picture. 
Whether a rigorous generalization of the van Cittert-Zernike theorem that includes time-dependent Doppler effects leads directly to the form used in this paper is an open theoretical question. 

We modify the temporal phase term by considering that the frequency $\nu$ of light from a moving part of the source is not constant but is Doppler-shifted by its LoS velocity $v_z$.
The frequency observed is given by the first-order Doppler effect:
\begin{equation}
\nu = \nu_0 \left(1 + \frac{v_z}{c}\right)
\end{equation}
where $\nu_0$ is the rest frequency. Substituting this $\nu$ into the spatio-temporal phase term $\exp{(-2\pi i\nu(t_i - t_j))}$ leads to
\begin{equation}
    \Gamma(u_i,v_i,t_i;u_j,v_j,t_j) = \iint_{\text{source}} I(x,y,\nu)\,
       e^{-2\pi i \phi(x,y)}\, dxdy
    \label{eq:Gamma_app2}
\end{equation}
with the phase
\begin{align}
    &\phi(x,y)=\nu_0 \left(1 + \frac{v_z}{c}\right)\left[(t_i-t_j)+\frac{u_i-u_j}{cd}x+\frac{v_i-v_j}{cd}y\right] \nonumber \\
    &=\nu_0(t_i-t_j)+\nu_0\frac{u_i-u_j}{cd}x +\nu_0\frac{v_i-v_j}{cd}x + \frac{\nu_0v_z}{c}(t_i-t_j)+\nu_0\frac{u_i-u_j}{c^2d}v_zx+\nu_0\frac{v_i-v_j}{c^2d}v_zx \nonumber \\
      &\approx \nu_0\frac{u_i-u_j}{cd}x + \nu_0\frac{v_i-v_j}{cd}x +\frac{\nu_0v_z}{c}(t_i-t_j)
\end{align}
Here we disregarded the second order terms as well as the constant phase offset $\nu_0(t_i-t_j)$.
Finally, expressing the remaining essential terms through the central wavelength at which we are observing the Doppler shifting emission or absorption line $\lambda_0=c/\nu_0$, we arrive at the phase used in Equation \ref{eq:visib}
\begin{equation}
      \phi(x,y)=\frac{1}{\lambda_0}\left[\frac{u_i-u_j}{ d}x + \frac{v_i-v_j}{ d}x +v_z(t_i-t_j)\right]
      \label{eq:phase}
\end{equation}
This provides the physical justification for the heuristic model used in the main body of this work to explore the observable consequences of spatio-temporal coherence.

\end{document}